# Scaling Behavior of Portevin-Le Chatelier Effect in Al-2.5%Mg Alloy

P. Barat[*], A. Sarkar, P. Mukherjee and S. K. Bandyopadhyay

Variable Energy Cyclotron Centre, 1/AF Bidhan Nagar,

Kolkata – 700 064, India

Abstract:

*The scaling behavior of the Portevin-Le Chatelier (PLC) effect was studied by deforming Al-2.5%Mg alloy for a wide range of strain rates. To reveal the exact scaling nature, the time series data of true stress vs. time, obtained during deformation, were analyzed by two complementary methods: the finite variance scaling method and the diffusion entropy analysis. From these analyses we could establish that in the entire span of strain rates, PLC effect showed Levy walk property.*

The study of the response of metals to applied force is primarily a subject of Mechanical Metallurgy, but the plastic instabilities observed in the stress-strain characteristics of metals have attracted attention to many people from various scientific disciplines. In the recent years plastic instabilities have been a subject of intensive investigations in the context of self-organisation in non-linear systems far from equilibrium. Among these instabilities, the Portevin-Le Chatelier (PLC) effect or Jerky flow in dilute alloys has been studied most extensively [1,2,3]. Typical of this effect is the complex behavior of discontinuous yielding or jerky flow in time and different types of strain localizations in space. The effect has continuously drawn attention due to its intriguing spatio-temporal dynamics. In this respect, the PLC effect falls into the class of complex non-linear driven systems exhibiting intermittent relaxation sequences, of which there are many examples [4-6]. In uniaxial loading with a constant imposed strain rate, the effect manifests itself as serrations in the stress vs. time (or strain) curves, associated with localized bands, static or propagating, of plastic deformation. The phenomenon is observed in many dilute alloys within a definite range of strain rates and temperatures. The physical origin of the PLC effect is known to arise from microstuctural process denoted by Dynamic Strain Aging (DSA), that is the dynamic interaction between mobile dislocations and mobile solute atoms [7-13]. Mobile dislocations, which are carriers of plastic strain, move jerkily between the obstacles provided by other dislocations. Solute atoms diffuse in the stress field generated by mobile dislocations, and further pin them while they are arrested at obstacles. This dynamic strain aging leads to a negative strain rate sensitivity of the flow stress within a certain range of applied strain rates and temperatures when mobile dislocations and solute atoms have comparable mobility [9,11,14,15]. Bands of localized deformation are then formed, in association with stress serrations. Close investigations of the PLC effect revealed the occurrence of different types of stress serrations. These serrations are well characterized in polycrystals, where they exhibit three main types of behavior: static, hopping and propagating, which are traditionally labeled as type C, B and A respectively [16-20]. Type C bands appear almost at random in the sample without propagating, type B bands exhibit an oscillatory or intermittent propagation, and type A bands, finally propagate continuously. Recent analysis [21,22] suggests that distinct dynamic features could be associated with each of these band types. At low strain rates static (type C) bands are associated with weak spatial interactions, consistent with randomness in their spatial distribution. In contrast at high strain rates, strong spatial correlations are associated with type A propagating bands, leading to self-organized critical (SOC) regime. At medium strain rates, partially relaxed spatial interactions lead to marginal spatial coupling linked to type B hopping bands. In this case, a chaotic regime was demonstrated.

---

[*] Corresponding author: pbarat@veccal.ernet.in

Due to a continuous effort of numerous researchers, there is now a reasonable understanding of the mechanisms and manifestations of the PLC effect. A review of this field can be found in [1,2]. The possibility of chaos in the stress drops of PLC effect was first predicted by G. Ananthakrishna *et.al* [23] and latter by V. Jeanclaude *et. al.* [ 24]. This prediction generated a new enthusiasm in this field. In last few years, many statistical and dynamical studies have been carried out on PLC effect. M. Lebyodkin *et. al.* have studied the spatio-temporal dynamics of the PLC effect in detail [25,26]. In one of their works they have also proposed the PLC effect to be a candidate for modeling of earthquake statistics [27]. Other researchers like M.S.Bharathi *et. al.* [22] and S. Kok *et. al.* [28] have studied the dynamical and chaotic behavior of the PLC effect.

Despite many dynamical studies, PLC effect remains an active area of research, with many important questions still open. This letter focuses on an alternate approach to establish the connection between the dislocation interactions and the stress fluctuations of the PLC effect. A link between these two phenomena is detected through a detailed scaling analysis of the time series data of the stress fluctuations during PLC effect in the different ranges of plastic instabilities in Al-2.5%Mg alloy. Scale invariance has been found to hold empirically for a number of complex systems [29], and the correct evaluation of the scaling exponents is of fundamental importance in assessing if universality classes exist [30].

Al-Mg alloys containing a nominal percentage of Mg are found to exhibit PLC effect at room temperature for a wide range of strain rates [18]. Tensile testing was conducted on flat specimens prepared from polycrystalline Al-2.5%Mg alloy. Specimens with gauge length, width and thickness of 25, 5 and 2.3 mm, respectively were tested in a servo controlled INSTRON (model 4482) machine. All the tests were carried out at room temperature (300K) and consequently there was only one control parameter, the applied strain rate. To monitor closely its influence on the dynamics of jerky flow, strain rate was varied from $7.56 \times 10^{-6}$ Sec$^{-1}$ to $1.92 \times 10^{-3}$ Sec$^{-1}$. The PLC effect was observed through out the range. The stress-time response was recorded electronically at periodic time intervals. Fig. 1 shows the observed PLC effect in a typical stress-strain curve for strain rate $3.85 \times 10^{-4}$ Sec$^{-1}$. The inset shows the magnified view of stress-time variation of a typical region in it. In the varied strain rate we could observe type A, B and C bands as reported [18,28].

To study the scaling behavior of the PLC effect we make use of two complementary scaling analysis methods: the finite variance scaling method (FVSM) and the diffusion entropy analysis (DEA) [31-35]. The need for using these two methods to analyze the scaling properties of a time series is to discriminate the stochastic nature of the data: Gaussian or Levy [33]. Recently, Scafetta *et al.* [36] had shown that to distinguish between fractal Gaussian intermittent noise and Levy-walk intermittent noise, the scaling results obtained using DEA should be compared with that obtained from FVSM.

These methods are based on the prescription that numbers in a time series $\{\xi_i\}$ are the fluctuations of a diffusion trajectory; see Refs. [32,37,38] for details. Therefore, we shift our attention from the time series $\{\xi_i\}$ to probability density function (pdf) $p(x,t)$ of the corresponding diffusion process. Here x denotes the variable collecting the fluctuations and is referred to as the diffusion variable. The scaling property of $p(x,t)$ takes the form

$$p(x,t) = \frac{1}{t^\delta} F\left(\frac{x}{t^\delta}\right) \quad (1)$$

In the FVSM one examines the scaling properties of the second moment of the diffusion process generated by a time series. One version of FVSM is the standard deviation analysis (SDA) [32], which is based on the evaluation of the standard deviation $D(t)$ of the variable x, and yields [29,32].

$$D(t) = \sqrt{\langle x^2;t \rangle - \langle x;t \rangle^2} \propto t^H \quad (2)$$

The exponent $H$ is interpreted as the scaling exponent and is usually called the Hurst exponent. It is evaluated from the gradient of the fitted straight line in the log-log plot of $D(t)$ against $t$.

DEA introduced recently by Scafetta *et al.* [31] focuses on the scaling exponent $\delta$ evaluated through the Shannon entropy $S(t)$ of the diffusion generated by the fluctuations $\{\xi_i\}$ of the time series using the pdf (1) [31,32]. Here, the pdf of the diffusion process, $p(x,t)$, is evaluated by means of the subtrajectories $x_n(t) = \sum_{i=0}^{t} \xi_{i+n}$ with $n=0,1,..$ Using Eq. (1) we arrive at the expression for $S(t)$ as

$$S(t) = -A + \delta \ln(t) \quad (A = \text{Constant}) \quad (3)$$

Eq. (3) indicates that in the case of a diffusion process with a scaling pdf, its entropy $S(t)$ increases linearly with $\ln(t)$. The scaling exponent $\delta$ is evaluated from the gradient of the fitted straight line in the linear-log plot of $S(t)$ against $t$.

Finally we compare H and $\delta$. For fractional Brownian motion the scaling exponent $\delta$ coincides with the Hurst exponent [32]. For random noise with finite variance, the diffusion distribution $p(x,t)$ will converge, to a Gaussian distribution with $H = \delta = 0.5$. If $H \neq \delta$ the scaling represents anomalous behavior. An interesting example of the anomalous diffusion is the case of Levy-walk, which is obtained by generalizing the central limit theorem [39]. In this particular kind of diffusion process the second moment is finite but the scaling exponents $H$ and $\delta$ are found to obey the relation

$$\delta = \frac{1}{3 - 2H} \qquad (4)$$

[32,38], instead of being equal.

Recently DEA and SDA have been applied to various social, meteorological, economical and biological time series data [36-38,40,41] to reveal the exact scaling nature. Here we take the initiative to apply these two methods for the time series data of the stress drop of the PLC effect in Al-2.5%Mg alloy to find the exact scaling. Fig. 2 and Fig. 3 show the plot of $D(t)$ and $S(t)$ against $t$ respectively calculated from stress vs. time data taken at $3.85 \times 10^{-4}$ Sec$^{-1}$ strain rate. These plots are fitted with Eqs. (2) and (3) respectively yielding the scaling exponents $H = 0.96$ and $\delta = 0.91$.

The scaling exponents $H$ and $\delta$ obtained from the time series data for different strain rates are listed in Table 1. The high values of the scaling exponents imply a strong persistence in the stress fluctuations. The values of $H$ and $\delta$ decrease marginally due to increase in strain rate. High strain rate demands higher average dislocation velocity and lesser waiting time, causing decrease in the additional activation enthalpy $\Delta G$ due to solute concentration accumulated at the glide dislocation segments. This enhances the probability of the thermally-activated dislocation glide, resulting more load drops in unit time. This causes $H$ and $\delta$ to decrease with increase in strain rate. Finally, we note that the standard deviation scaling exponents ($H$) are larger than the corresponding diffusion entropy scaling exponents ($\delta$) and seen to fulfill the Levy-walk diffusion relation (Eq. (4)) within the accuracy of our statistical analysis as shown in the column 4 of Table 1.

It is evident from different studies [3,42] that the physics governing the PLC effect necessarily localizes the deformation in the form of bands at all plastic strain level after the critical strain at which PLC effect initiates. Metals deform through the generation and propagation of dislocations, in the sub-micron scale. The key to the PLC effect lies ultimately bundled in the dynamics that connect the microscopic world of dislocations to the macroscopic regime of the bands. Deformation bands travel with constant velocity at constant stress [43]. If the pulling speed is increased, the band velocity changes proportionally. Initial motion of the band is accompanied by a sudden drop in stress. For low strain rate, this drop is sufficient to stop the band. At high strain rates, the stress drops are relatively small in magnitude and no longer stop the band completely and it travels from hopping to constant velocity with the increase in the strain rate. Depending on temperature and strain rate conditions these bands may or may not be correlated in space. This correlation arises due to the long range elastic interactions. In the earlier works [21,22] it was shown that the dynamics of the PLC bands at low and medium strain rates are chaotic while at high strain rate it showed SOC. But our results from DEA and SDA (Table 1) clearly suggest that the scaling behavior of the overall dynamics of the PLC effect at all strain rates follow Levy-walk property. The strongly correlated glide of macroscopic dislocation groups in the band, the long range elastic interactions among the dislocations and the DSA are the basic ingredients for the dynamics of the PLC bands. The variations of the degree of these three attributes manifest different observable macroscopic dynamic characteristics of the bands. Identicality of the basic dynamical features responsible for the PLC bands, made them to scale uniformly.

In conclusion, we have studied the PLC effect from a new perspective and evaluated the exact scaling behavior of the PLC effect in Al-2.5%Mg alloy using two complementary scaling analysis methods: DEA and SDA. The analyses were performed in a wide range of strain rates where different types of deformation bands are observed. The relation of the two scaling exponents in each strain rate obtained through our analysis clearly suggests that the stress fluctuations occurring due to dislocation flow (PLC effect) in Al-2.5%Mg alloy inherit a Levy-walk memory component and the scaling behavior is independent of strain rate.

TABLE 1. SDA and DEA scaling exponents obtained from the Stress vs. Time data of Al-2.5%Mg alloy during tensile deformation for different strain rates.

| Strain rate (Sec$^{-1}$) | SDA exponent ($H$) (Maximum error=±0.02) | DEA exponent ($\delta$) (Maximum error=±0.02) | $\left[\left(\delta - \dfrac{1}{3-2H}\right)/\delta\right] \times 100$ |
|---|---|---|---|
| 7.56×10$^{-6}$ | 0.98 | 0.94 | 2.29 |
| 1.99×10$^{-5}$ | 0.98 | 0.93 | 3.39 |
| 3.98×10$^{-5}$ | 0.97 | 0.92 | 2.54 |
| 7.99×10$^{-5}$ | 0.96 | 0.92 | 0.64 |
| 1.55×10$^{-4}$ | 0.96 | 0.92 | 0.64 |
| 3.85×10$^{-4}$ | 0.96 | 0.91 | 1.75 |
| 5.88×10$^{-4}$ | 0.96 | 0.91 | 1.75 |
| 7.42×10$^{-4}$ | 0.96 | 0.91 | 1.75 |
| 1.21×10$^{-3}$ | 0.95 | 0.90 | 1.01 |
| 1.59×10$^{-3}$ | 0.94 | 0.89 | 0.32 |
| 1.92×10$^{-3}$ | 0.94 | 0.88 | 1.46 |

**Figure Captions:**

FIG. 1. True Stress vs. True Strain curve of Al-2.5%Mg alloy at a strain rate of 3.85×10$^{-4}$ Sec$^{-1}$. The inset shows a typical region of the curve in the Stress-Time plot.

FIG. 2. SDA of the Stress vs. Time data obtained from Al-2.5%Mg alloy during tensile deformation at a strain rate of 3.85×10$^{-4}$ Sec$^{-1}$.

FIG. 3. DEA of the Stress vs. Time data obtained from Al-2.5%Mg alloy during tensile deformation at a strain rate of 3.85×10$^{-4}$ Sec$^{-1}$.

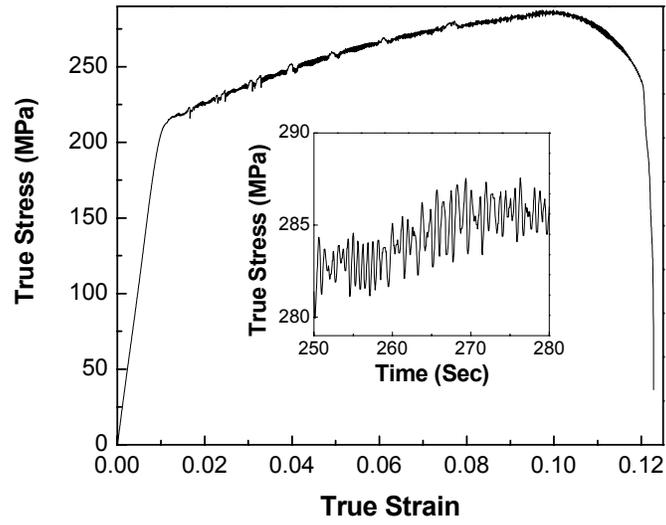

FIG. 1., P.Barat, Physical Review Letters

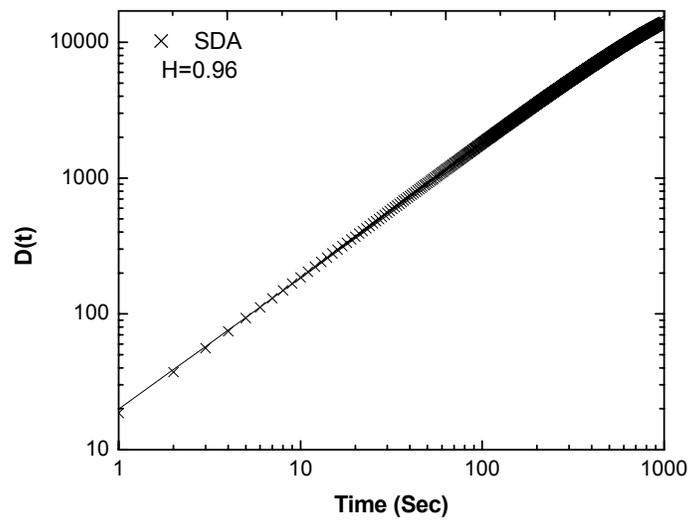

FIG. 2, P. Barat, Physical Review Letters.

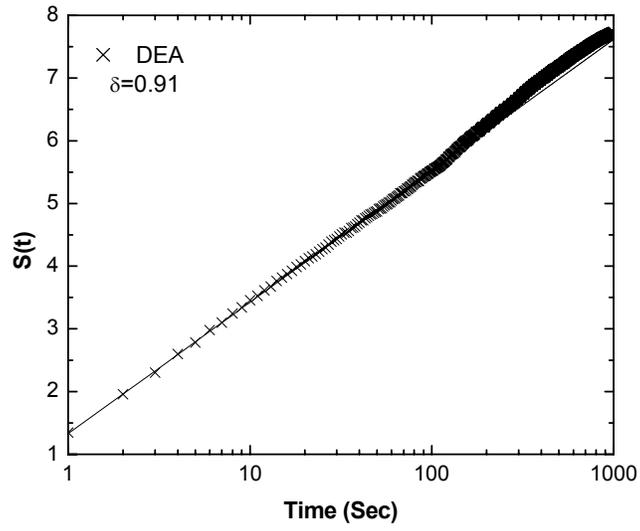

FIG. 3., P.Barat, Physical Review Letters.